# A Circuit-based Model for the Interpretation of Perfect Metamaterial Absorbers

Filippo Costa, *Member, IEEE*, Simone Genovesi, *Member, IEEE*, Agostino Monorchio, *Fellow, IEEE*, Giuliano Manara, *Fellow, IEEE*

*Abstract* — A popular absorbing structure, often referred to as Perfect Metamaterial Absorber, comprising metallic periodic pattern over a thin low-loss grounded substrate is studied by resorting to an efficient transmission line model. This approach allows the derivation of simple and reliable closed formulas describing the absorption mechanism of the subwavelength structure. The analytic form of the real part of the input impedance is explicitly derived in order to explain why moderate losses of the substrate is sufficient to achieve matching with free space, that is, perfect absorption. The effect of the constituent parameters for tuning the working frequency and tailoring the absorption bandwidth is addressed. It is also shown that the choice of highly capacitive coupled elements allows obtaining the largest possible bandwidth whereas a highly frequency selective design is achieved with low capacitive elements like a cross array. Finally, the angular stability of the absorbing structure is investigated.

## I. INTRODUCTION

Electromagnetic absorbers have aroused outstanding interest due to their range of application which spans from the microwave to optical frequency regime passing through THz spectrum [1]. At microwaves they are employed as electrically thin layers to reduce the radar signature of targets [2]-[6], for power imaging purposes [7], to improve the electromagnetic compatibility of electronic devices [8], [9] or even as Chipless Radio Frequency Identification tags [10]. In the THz range they are used in photodetectors or microbolometers [11], [12] and phase modulators [13]. Recently, selective absorbing structures have been proposed in optical regime as thermal emitters matched with the bandgap of solar cells to improve the efficiency of thermophotovoltaic systems [14]-[16]. A large number of thin absorber designs based on engineered structures has been presented in the last few years [11]-[20]. Among these designs, the most successful configuration is formed by a periodic surface printed on a grounded dielectric layer. Such periodic structure was already known in the microwave regime with the name of High-Impedance Surface (HIS) [21], [22]. The resonant subwavelength structure is able to perform perfect absorption at a single frequency or within a wideband frequency if a suitable amount of losses is introduced. Losses can be introduced in different ways depending on the addressed frequency region. At microwaves, lumped resistors or resistive inks are employed to synthesize both narrowband and wideband absorbers. Alternatively, a perfect absorption can be achieved within a narrow frequency band by simply exploiting the intrinsic loss factor of commercial substrates. The former structure is mainly based on resistive losses and it has been extensively studied in microwave regime [23]-[29] and interpreted in terms of circuital model [2], [3], [4], [30], [31]. The latter configuration is instead frequently referred to as Perfect Metamaterial Absorber [17] and it is usually explained by recurring to complex effective medium parameters $\varepsilon$ and $\mu$ even if more classical theories based on multiple reflections interactions have been recently explored to explain the physical mechanisms of this subwavelength structure [32], [33].

Here, a simple equivalent circuit model of the metamaterial absorber, including the effect of metal and dielectric losses, is exploited to study the dissipative mechanisms of the absorbing structure. We derive closed form expressions which allow highlighting the effect of ohmic or dielectric losses and the role played by the geometrical parameters of the absorbers. Once derived the closed-form expressions, it is easy to observe the influence of the element shape, the substrate thickness and the dielectric tangent loss on the reflection amplitude reduction. It will be pointed out that the loss mechanism is due only to the electrical absorption since there are no magnetic phenomena responsible for energy dissipation in the thin absorber. The derived expressions provide a clear model for the interactions causing the reflection losses.

The paper is organized as follows: in the next Section, the absorption properties of high-impedance surfaces are represented from a circuital point of view. Sect. III and Sect. IV are dedicated to the calculation of the input impedance of a grounded lossy substrate and of the impedance of a periodic array printed on a lossy dielectric. In Sect. V the real and the imaginary parts of the input impedance of the entire resonant structure are calculated by merging the results achieved in the two previous sections. Section VI presents a discussion on the influence of electrical and geometrical parameters on the absorption properties of the metamaterial absorber. Some examples aimed to verify the validity of the derived closed-form expressions are also presented. Finally, in section VII, the effect of oblique incidence is assessed.

## II. ANALYTICAL DERIVATION OF ABSORPTION PROPERTIES OF THE HIGH-IMPEDANCE SURFACE

The layout of the investigated structure and its equivalent circuit are reported in Fig. 1. The structure is formed by a metallic Frequency Selective Surface (FSS) printed on a lossy grounded dielectric substrate and it is basically a subwavelength resonant cavity characterized by an input impedance approaching to infinite and a reflection phase crossing zero at the resonance [21]. The amount of power absorbed by the resonant cavity at the resonance is determined by the value of the real part of the input impedance. As a matter of fact, the magnitude of the reflection coefficient of the periodic structure reads:

$$|\Gamma| = \sqrt{\frac{(\operatorname{Re}\{Z_H\}-\zeta_0)^2 + (\operatorname{Im}\{Z_H\})^2}{(\operatorname{Re}\{Z_H\}+\zeta_0)^2 + (\operatorname{Im}\{Z_H\})^2}} \quad , \quad (1)$$

where $Z_H$ represents the input impedance of the absorbing HIS structure and $\zeta_0$ is the characteristic impedance of free space.
For the ideal lossless structure, the real part of the input impedance is zero and the reflection coefficient magnitude always equals the unity. The input impedance of an actual HIS structure realized with lossy substrate is instead characterized by a very high real part and by the typical smoothed transition through zero of the imaginary part [3]. A very high real part leads to a limited amount of reflection losses. As the real part of the input impedance $Z_H$ decreases down to the free space impedance, the HIS structure performs a progressive absorption of the incoming signal.

An explicit expression of the real part of the input impedance of the high-impedance surface $Z_H$ is obtained by separately deriving and then combining the expressions of the grounded substrate impedance, $Z_d$, and the FSS impedance, $Z_{FSS}$ on the basis of simple approximations.

## III. COMPLEX INPUT IMPEDANCE OF A GROUNDED LOSSY DIELECTRIC SLAB

The input impedance of a grounded dielectric slab $Z_d$ behaves as an inductor until its thickness is lower than a quarter wavelength. At normal incidence, its analytical expression is the following:

$$Z_d = j\frac{\zeta_0}{\sqrt{\varepsilon_r' + j\varepsilon_r''}} \tan\left(k_0\sqrt{\varepsilon_r' + j\varepsilon_r''}\, d\right) \quad (2)$$

where $\zeta_0$ is the characteristic impedance of free space; $k_0$ is the free space propagation constant and $d$ is the thickness of the dielectric substrate. Assuming $\varepsilon_r' \gg \varepsilon_r''$, the real and the imaginary part (named $A$ and $B$ respectively, for convenience) of the input impedance can be expressed as follows [34]:

$$\operatorname{Re}\{Z_d\} = A \cong \frac{\zeta_0}{\sqrt{\varepsilon_r'}}\left[\frac{\varepsilon_r''}{2\varepsilon_r'} tg\left(k_0 d\sqrt{\varepsilon_r'}\right) - \left(k_0 d \frac{\varepsilon_r''}{2\sqrt{\varepsilon_r'}}\right)\left(1 + tg^2\left(k_0 d\sqrt{\varepsilon_r'}\right)\right)\right] \quad (3)$$

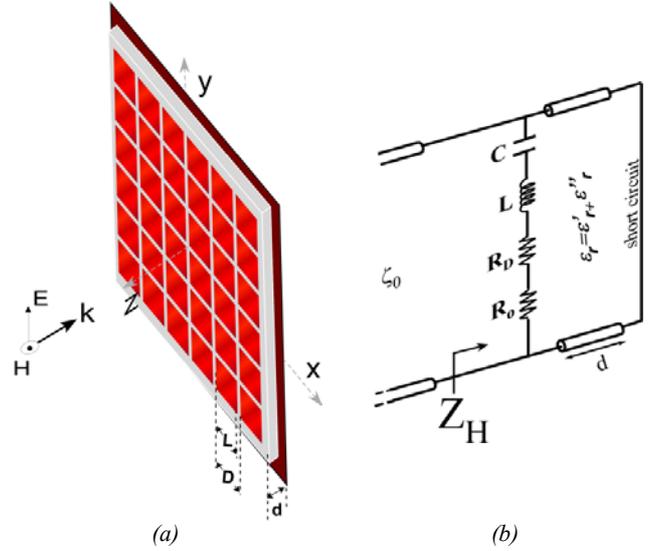

*(a)*                                *(b)*

Figure 1. 3D sketch of the analyzed structure *(a)* and its equivalent circuit *(b)*.

$$\operatorname{Im}\{Z_d\} = B \cong \frac{\zeta_0}{\sqrt{\varepsilon_r'}}\left[tg\left(k_0 d\sqrt{\varepsilon_r'}\right)\right] \quad . \quad (4)$$

The real part of the input impedance depends both on the real and imaginary part of the dielectric permittivity while the imaginary part of $Z_d$ is almost equal to the lossless case.

## IV. COMPLEX IMPEDANCE OF A FREQUENCY SELECTIVE SURFACE PRINTED ON A LOSSY SUBSTRATE

The impedance of a lossless FSS is capacitive before its proper resonance and becomes inductive after it. The impedance can be represented through a series LC circuit or more simply by a single capacitor if the inductive component is low (e. g. patch element):

$$Z_{FSS}^{lossless} = jX = \left(1 - \omega^2 L C_0\right)/\left(j\omega C_0\right) \quad . \quad (5)$$

If the FSS is printed on a lossy dielectric substrate, its capacitance can be computed by multiplying the unloaded capacitor by the effective dielectric permittivity due the surrounding dielectrics [35]. If the hypothesis of a sufficiently thick substrate is verified (thicker than 0.3 $D$, where $D$ is FSS periodicity [35], the effective permittivity, $\varepsilon_{reff}$, simply corresponds to the average between the relative permittivity of the substrate ( $\varepsilon_r = \varepsilon_r' + j\varepsilon_r''$ ) and the relative permittivity of free space [34]. Therefore, the loaded capacitance reads:

$$C = C_0\left[\frac{(\varepsilon_r'+1)}{2} + j\frac{\varepsilon_r''}{2}\right] \quad \text{if } (d > 0.3D) \quad . \quad (6)$$

The imaginary part of the capacitor in (6) leads to a resistive component in the FSS impedance which takes into account the effect of the lossy substrate in the proximity of a metallic array. The capacitor formed between the adjacent elements has a loss component since electric field lines are concentrated in a lossy medium. Such loss component is readily represented by a resistor in parallel with the lossless capacitor but in our formulation it is more convenient to transform the shunt connection in a series of a capacitor and a resistor $R_D$ [34]:

$$R_D \simeq \frac{-\varepsilon_r''}{\omega C_0 \left(\varepsilon_r' + 1\right)} \qquad . \qquad (7)$$

Ohmic losses can be taken into account by an additional resistor $R_O$ connected in series with the aforementioned dielectric resistor. $R_O$ can be evaluated by weighting the classical expression of the surface resistance of metals with the ratio between metalized area and the square of element periodicity [3]:

$$R_o \approx \left(\frac{D}{L}\right)^2 \frac{1}{\delta\sigma} \qquad . \qquad (8)$$

Ohmic losses can be neglected in microwave range since the resistor in (8) is generally one or two orders of magnitude lower than the dielectric resistor (7). Conversely, if the metal is replaced by a resistive paint, the resistor assumes considerably higher values than the dielectric resistor [3]. Ohmic losses, which come from the currents flowing on an imperfect conductor, are instead increasingly important as the working frequency raises. In THz range, the ohmic resistor is comparable with the dielectric one while in optical regime ohmic losses dominate [36].

Since under the hypothesis of small losses in the substrate, the imaginary part of the FSS impedance is almost equal to the lossless case, the total FSS impedance is the following:

$$Z_{FSS}^{lossy} \simeq R_0 + R_D + jX \qquad (9)$$

The calculation of the unloaded capacitance can be accomplished by retrieving the reflection coefficient of a full-wave simulation [37]. Alternatively, in case of a patch FSS, it can be calculated through the closed-form expression available in [35]. As the substrate thickness is reduced (which is the case of thin metamaterial absorbers) the influence of higher-order (evanescent) Floquet modes reflected by the ground plane must be taken into account by adequately correcting the capacitance and the inductance values. In particular, the value of the capacitor increases exponentially as the spacer thickness is reduced below $0.3\,D$ [35], [37]. The influence of the evanescent modes can be taken into account by the following substitution [35]:

$$C_0^{thin} = C_0 - \frac{2D\varepsilon_0}{\pi}\log\left(1 - e^{-\frac{4\pi d}{D}}\right), \qquad (10)$$

where $d$ represents the thickness of the dielectric substrate. The FSS series resistor $R_D$ is inversely proportional to the FSS capacitance. The behavior of the total FSS resistance is reported in Fig. 2 as a function of frequency for three different substrate thicknesses below the aforementioned limit (d<0.3D). In this case, the metallic periodic surface is composed by an array of patches made of gold characterized by a side length of 14/16 $D$, where $D$, which represents the repetition period, is equal to 20 μm. In the same figure the capacitance of the patch array as a function of the substrate thickness and periodicity ratio is also shown. The result obtained by retrieving full-wave data is compared with the one obtained with relation (10). It is evident that a reduction of the substrate thickness below $0.3\,D$ determines an exponential growth of the lumped capacitor which is physically ascribed to the capacitance formed from the metallic array and the ground plane. This capacitance is negligible with respect to the capacitance formed between adjacent patches if the substrate is thicker than $0.3\,D$.

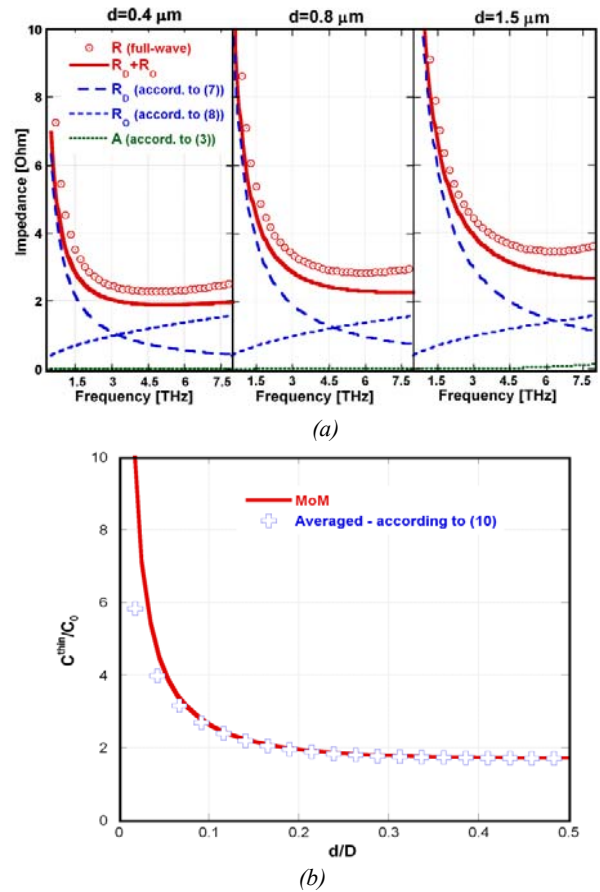

Figure 2 – Input impedance of the absorber as a function of the FSS capacitance. (a) Behavior of the total FSS resistor $R$ and its terms $R_O$ and $R_D$ as a function of frequency. The real part of the substrate input impedance $A$ is also reported for comparison. (b) Comparison between the capacitance of a patch array computed according to the averaged theory (relation (10)) and with a MoM simulation. Parameters: D = 20 μm, L = 15μm.

It turns out that the resistor $R_D$ decreases when the substrate thickness is lower 0.3D since the FSS capacitance $C_0$ increases (see relation (7)). The ohmic resistor does not vary with the substrate thickness according to the model. In Fig. 2 the sum of the analytically computed resistors (R=$R_O$+$R_D$) is compared with the total resistor retrieved from full-wave MoM simulations. The above described trend is correctly verified since also the full-wave resistor decreases as the substrate thickness is lower than 0.3 D. The small deviations are essentially due to the fact the model is of first order and it is valid up to the first resonance. The total resistance of the real structure tends to increase again as the second resonance is approached. This behavior could be modeled by introducing additional lumped elements in the FSS impedance but this complication would not allow obtaining the simple analytical relations here derived. The loss component of the grounded substrate input impedance, named $A$ in the figure, is negligible with respect to the FSS resistor in correspondence of the main resonance. This term becomes relevant as the frequency increases and, for this reason, it may become the main loss component of the structure in correspondence of higher order resonances [10].

## V. INPUT IMPEDANCE OF THE HIS STRUCTURE AS A FUNCTION OF SUBSTRATE LOSS

In the previous sections, the impedance of the periodic printed surface and the input impedance of the lossy grounded substrate have been expressed as complex numbers characterized by a real and an imaginary part. The input impedance of the HIS structure $Z_H$ is equal to the parallel connection between the two complex impedances $Z_{FSS}$ and $Z_d$. After some simple algebra, the real part of the input impedance $Z_H$ can be expressed as follows:

$$\text{Re}\{Z_H\} = \frac{(AR-BX)(A+R)+(BR-XA)(B+X)}{(A+R)^2+(B+X)^2} \quad (11)$$

If $A = 0$ the formulation collapses to the case of thin resistive HIS absorbers [3] in which only ohmic losses are considered. The resonance condition of the structure is fulfilled if the imaginary part of the input impedance $Z_H$ equals zero (that is $X \approx -B$ if $B \gg A$ and $B \gg R$) [34], which means that the resonance of the structure is almost unchanged with respect to the lossless case. Assuming verified the equality $X = -B$, the real part of the input impedance of the structure at the resonance is derived from (11):

$$\text{Re}\{Z_H^{res}\} \cong \frac{B^2}{(A+R)} \quad (12)$$

By replacing the relations (7), (3), (4) in (12), the real part of the input impedance at the resonance can be explicitly written. Assuming that the resistance of the FSS, $R_O+R_D$, is much higher than the real part of the input impedance of the grounded dielectric slab, $A$, in correspondence of the first resonance of the structure (see Fig. 2 and also the analytical demonstration in appendix III of [34]), the relation (12) can be further simplified up to the first resonance to obtain the real part of the input impedance of the absorber:

$$\text{Re}\{Z_H^{res}\} = \frac{\dfrac{\zeta_0^2}{\varepsilon_r'}\left[tg^2\left(k_0 d\sqrt{\varepsilon_r'}\right)\right]}{\left(\dfrac{D}{L}\right)^2 \dfrac{1}{\delta\sigma} + \dfrac{-2\varepsilon_r''}{\omega_0 C_0 \left(\varepsilon_r'+1\right)^2}} \quad (13)$$

where $\omega_0$ is the first resonance frequency. The expression of the real part of $Z_H$ in (13) contains all the degrees of freedom of the metamaterial absorber: it is a function of the FSS capacitance, of the electrical substrate thickness and of the real and imaginary part of the dielectric permittivity. In the next section, the effects of these parameters on the absorption mechanisms of the absorber will be discussed in detail.

## VI. DISCUSSION

### a) Substrate thickness

The use of a thin substrate tends to gradually lower the real part of the input impedance with a consequent variation of the reflection coefficient magnitude. As the input impedance becomes equal to the free space impedance we achieve perfect absorption (or perfect matching) but a further reduction of the substrate thickness leads to a reduction of the absorption. The latter condition can be also observed by looking at the phase of the reflection coefficient since, in case of perfect matching with the free space, it is characterized by an abrupt phase jump [38].

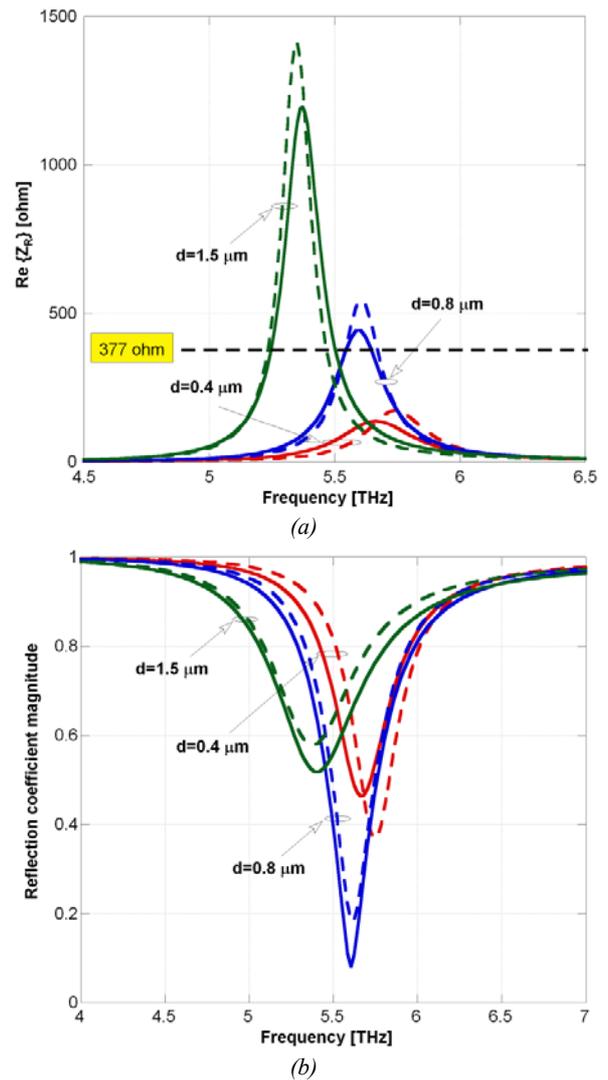

Figure 3 – Analysis of the performance of the absorber as a function of the substrate thickness. Solid lines: MoM simulations; Dashed lines: TL model. *(a)* Real part of the input impedance of the absorber for three different substrate thicknesses. *(b)* Reflection coefficient of a patch array made of gold printed on a Mylar substrate with three different thicknesses. Parameters: $D = 20$ μm, $L = 15$ μm.

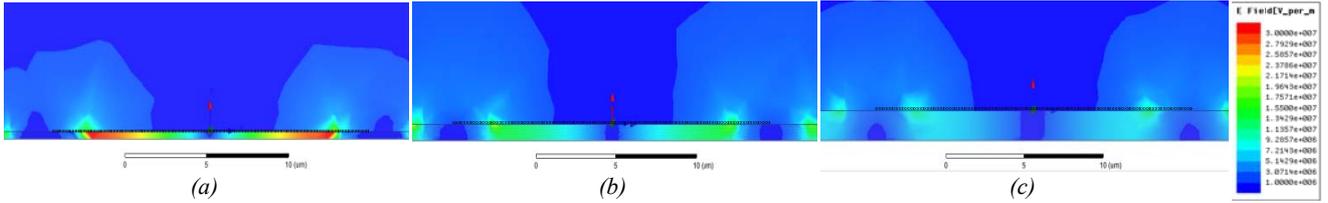

Figure 4 – Electric field distribution of the high-impedance surface with three different substrate thickness: *(a)* 0.4 μm, *(b)* 0.8 μm, *(c)* 1.5 μm.

In order to clarify these concepts, we have simulated an array of patches made of gold printed on a Mylar substrate (εr = 2.89, tg δ = 0.02) in the THz range. The reflection coefficient of the structure for three different thicknesses of the substrate is reported in Fig. 3 together with the input impedance of the absorber. In Fig.4 the electric field distribution for the three analyzed cases is reported. As the thickness is increased, a moderate shift toward lower frequencies is achieved because of the increase of the substrate inductance. The electrical increase of the thickness is lower than the physical one because of the shift of the resonance towards larger wavelengths. The shift is also mitigated by the reduction of the FSS capacitance as shown in Fig. 2. As the thickness of the substrate is reduced a high confinement of the electric field below the patches is observed and an equal value of substrate loss leads to a different amount of dissipated power because of a different matching with free space. For the analyzed configuration, an almost perfect matching is achieved for a thickness of 0.8 μm while a lower and a thicker substrate leads to a high mismatch. By observing the input impedance of the absorber, it is evident that the 0.8 μm substrate case leads to a real part of the input impedance very close to the value of the free space impedance (i.e. 377 Ω).

### b) Substrate losses

The imaginary part of the substrate permittivity appears at the denominator of (13). As expected, the increase of the loss component in the substrate causes a reduction of the real part of the input impedance.

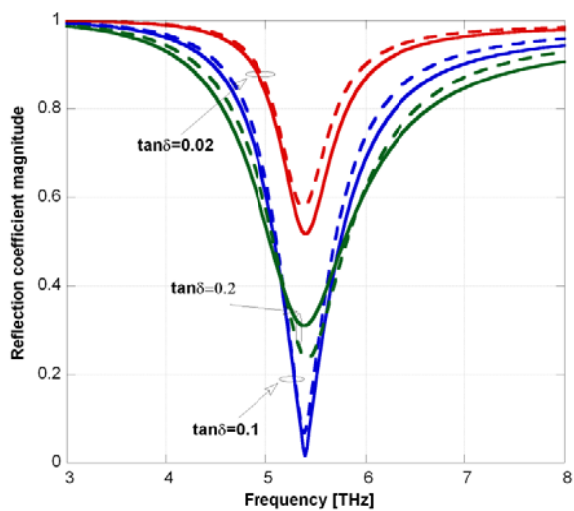

Figure 5 – Reflection coefficient of a patch array made of gold printed on a 1.5 μm thick substrate as a function of the substrate loss component.

Analogously to the previous case, there exists an optimal tangent loss leading to a perfect matching while if a lower or higher amount of loss is chosen an absorbance reduction is obtained. In Fig. 5 the reflection coefficient of an array of patches made of gold printed on a 1.5 μm substrate is reported for three different values of tangent loss. As it is evident, the optimal tangent loss for achieving perfect absorption with the chosen dielectric thickness is equal to 0.1. A lower amount of loss leads to an input impedance higher than the free space impedance while a too lossy substrate would reduce the absorbance since the real part of the input impedance becomes smaller than the free space impedance.

### c) Unit cell shape

The choice of the unit cell of the periodic FSS layer is a key task in the design of a thin metamaterial absorber. In Fig. 6 the electric field surface distributions of three HISs with different FSS elements printed on the same substrate are illustrated.

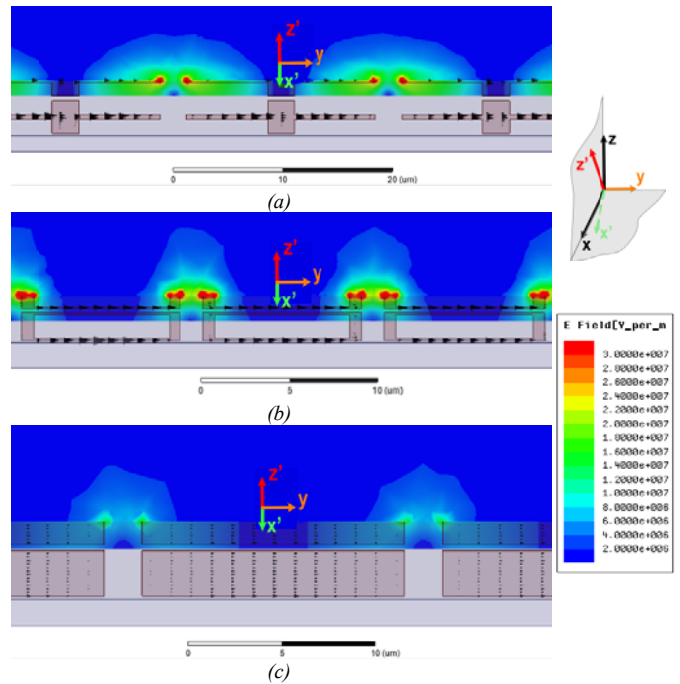

Figure 6 – Electric field and surface current distribution for three different high-impedance surfaces comprising different unit cells but with the same substrate thickness and resonating at the same frequency: *(a)* cross, *(b)* ring, *(c)* patch. The ground plane is located on *xy* plane.

The electric field is characterized by a planar and an orthogonal component [30]. If the substrate is thicker than 0.3*D*, the planar component of the electric field is predominant on the orthogonal component but, if the thickness is lower than 0.3 *D*, the orthogonal component of the electric field becomes predominant. The latter phenomenon is represented in the circuit model by the exponential growth of the capacitance shown in section II. A close look to (13) reveals that the use of a highly capacitive element contributes to enhance the value of the real part of the input impedance. Keeping the substrate thickness fixed, a coupled element (*e.g.* patch with small gap) requires higher substrate losses than a low capacitive element, like a cross array, to achieve perfect absorption. The reason of this behavior is that the electric fields are concentrated between the adjacent edges and behind the metallic regions. As a consequence the field intensity is higher on thin element types, like crosses or square loops and a lower amount of substrate loss leads to a total absorption.

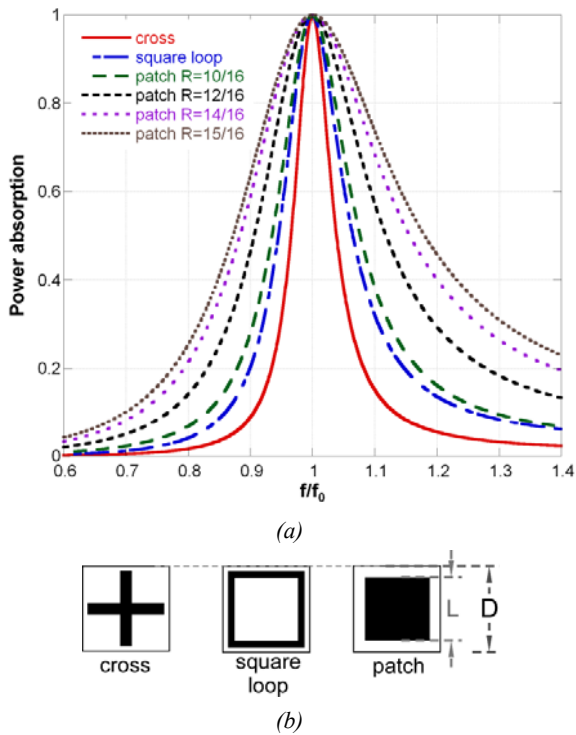

Figure 7 – Analysis of the power absorbed by the metamaterial depending on the printed pattern. (a) Power absorbed by metamaterials absorbers comprising different unit cell shapes (b) and the same substrate thickness. *R = L/D*.

In addition, according to (8), the elements with a small filling factor of the unit cell are characterized by high ohmic losses due to high surface currents. Patch type configuration allows minimizing the ohmic losses and maximize the dielectric ones but, as it has been already pointed out, it requires a high substrate loss component (tangent loss) or an adequately thin substrate to perform perfect absorption.

A highly capacitive element performs the absorption over the widest possible band for a given substrate thickness [3], [22]. The bandwidth of the absorber can be clearly enlarged also by increasing the thickness of the substrate [22], but the loss reduction due to the higher substrate has to be compensated by increasing the loss component of the substrate or by changing the FSS unit cell.

The aforementioned arguments are readily verified by analyzing the absorption properties of different FSS configurations dimensioned to absorb at the same frequency. The substrate thickness and permittivity are kept fixed to 1.5 μm and 2.89 respectively. The power absorbed by the metamaterial absorber composed by six different unit cells is reported in Fig. 7. As predicted, the most coupled patch element allows the absorption over the widest bandwidth since its capacitance is the highest one. On the contrary, the use of a low capacitive element as a cross array leads to the most selective absorption profile. In general, the capacitance of an FSS element grows up as the coupling area is maximized and the distance between the adjacent metal conductors is reduced. Moreover, the inductance of a printed structure slows down as the element is made larger or longer keeping fixed the substrate parameters. The values of the lumped parameters computed for the analyzed elements are reported in Table 1.

Table 1 – Geometrical and electrical properties of the analyzed absorbers. The -5 dB percentage bandwidth is also reported.

|  | C [fF] | L [pH] | D [μm] | $\varepsilon_r''$ | **BW%** |
|---|---|---|---|---|---|
| Cross | 0.0943 | 7.054 | 19.74 | 0.04 | 4.9 |
| Square loop1 | 0.157 | 3.374 | 10.39 | 0.042 | 8 |
| Square loop2 | 0.199 | 2.368 | 11.55 | 0.16 | 10 |
| Square loop3 | 0.238 | 1.695 | 12.75 | 0.24 | 12.4 |
| Patch R=10/16 | 0.187 | 2.71 | 24.5 | 0.17 | 9.8 |
| Patch R=12/16 | 0.265 | 1.359 | 20 | 0.289 | 14.7 |
| Patch R=14/16 | 0.334 | 0.665 | 16.2 | 0.39 | 18.3 |
| Patch R=15/16 | 0.358 | 0.445 | 14.02 | 0.47 | 20.5 |

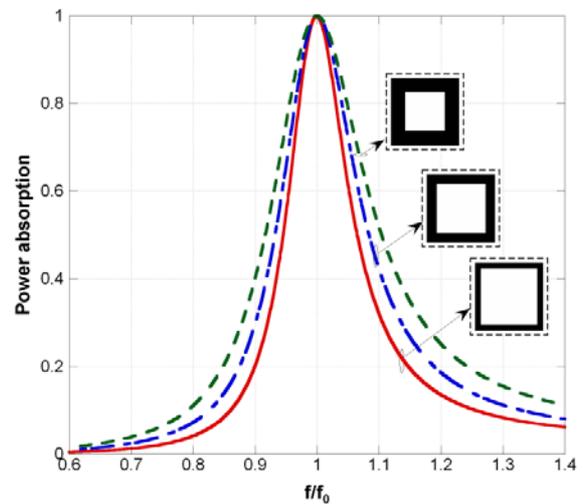

Figure 8 – Analysis of the power absorbed as a function of the width of the ring unit cell.

In Fig. 7, the thickness of the cross element is equal to 1/8 of its periodicity and the thickness of the loop is equal to 1/16 of its periodicity. The bandwidth can be modulated by adjusting the thickness of the element as apparent from the patch case. In Fig. 8 the power absorption of three different loops is also reported. As the loop is made wider its inductance is lowered. In order to achieve the resonance at the same frequency, the periodicity need to be enhanced leading to an increase of its equivalent capacitance. The choice of a thicker loop allows widening the operating bandwidth but, as already remarked, In conclusion, both the bandwidth and the absorbed/emitted power can be tailored for the specific requirements by tuning all the degrees of freedom of the subwavelength structure.

## VII. ANGULAR STABILITY

The name "perfect absorber" associated to this simple structure comes essentially from the high angular stability of the absorption profile. The angular stability performance mainly depends on the reduced thickness which characterizes the absorbing structure. The problem can be addressed analytically with the same approach followed at normal incidence. The input impedance of the absorber can be again computed according to relation (12) but the terms present in this relation can vary at oblique incidence in different manners for TE and TM polarization. The real part of the FSS impedance, as previously remarked, is composed by a series of two resistors, i.e. a ohmic resistor, $R_O$, and a dielectric resistor, $R_D$. The ohmic resistor can be assumed angle-independent whereas the dielectric resistor varies according to the angular dependence of the capacitance. For a patch array the capacitance is angular dependent for TE polarization [39] (it decreases as the incident angle increase) while the cross array capacitance is almost angle-independent [40]. The expression of the grounded substrate input impedance at oblique incidence reads:

$$Z_d = jZ_m^{TE,TM} \tan\left(k_0 d \sqrt{\varepsilon_r' + j\varepsilon_r'' - \sin^2(\vartheta)}\right), \quad (14)$$

where $Z_m^{TE} = (\omega\mu_r\mu_0)/\beta$; $Z_m^{TM} = \beta/(\omega\varepsilon_r\varepsilon_0)$ are the characteristic impedances of the slab for TE and TM polarization, $\beta = k_0\sqrt{\varepsilon_r - \sin^2(\vartheta)}$ is the propagation constant along the normal unit of the slab and $\theta$ is the incidence angle of the incoming wave with respect to the normal. It is possible to calculate in a closed-form the expression of the terms $A$ and $B$ for TE and TM incidence [34]. Anyway, as previously pointed out, metamaterial absorbers are necessarily made with very thin dielectric slabs (on the order of $\lambda/30$ or even thinner) for achieving perfect matching with free space impedance. The input impedance of a thin grounded substrate (d<< $\lambda$) as a function of the incident angle is well approximated by the following relations:

$$\begin{aligned} B^{TE} &\approx j\omega\mu_0 d \\ B^{TM} &\approx j\frac{\left(\varepsilon_r' - \sin^2(\vartheta)\right)}{\varepsilon_r'}\omega\mu_0 d \end{aligned} \quad . \quad (15)$$

The use of a thin dielectric substrate limits implicitly the effect of the incidence angle. As an example, the absorption profile of a cross array on top of a 1.5 µm thick mylar grounded substrate is reported in Fig. 9. It results from (15) that the reactance of the input impedance for TE polarization is almost unchanged with respect to the normal incidence one but the increase of $\theta$ leads to a decrease of the imaginary part of the input impedance for TM polarization.

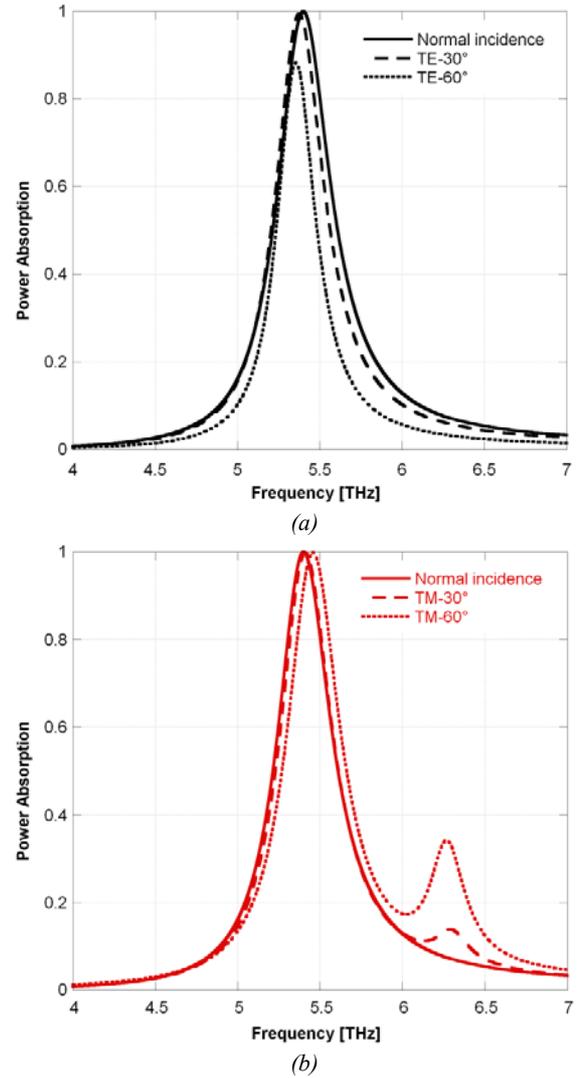

Figure 9 – Power absorbed by a cross shaped metamaterial absorber at TE (a) and TM (b) oblique incidence. Parameters: d=19.74 µm, $D$ = 20 µm, L = 15µm.

In the computation of the reflection coefficient, it has to be taken into account that the TM free space impedance drops as the incident angle increases ($Z_0^{TM} = Z_0 \cos(\vartheta)$) (the opposite is valid for TE incidence since $Z_0^{TE} = Z_0/\cos(\vartheta)$). The real part of the input impedance of the absorber for TM polarization is proportional to the square of the imaginary part of the input impedance of the grounded substrate $B$ (see (12)). Since $B$, according to (15), decreases as the incidence angle increases, a good absorption can be guaranteed for very wide incident angles by properly choosing a dielectric permittivity of the substrate able to compensate angular variation. In particular the following identity should be verified:

$$\left[\left(\varepsilon_r' - \sin^2(\vartheta)\right)/\varepsilon_r'\right]^2 \approx \cos(\vartheta) \quad . \quad (16)$$

It is interesting to observe that the choice of very high dielectric constant would limit the angular dependence of the relation in (15) but a non-optimal absorption would be obtained for very oblique incident angles. A wide incident angle perfect absorption for TE polarization is instead more difficult to obtain. The absorption at 60°, for instance, is around 90% since the free space impedance is nearly $2Z_0$ while the input impedance of the structure, which is almost angle independent, is still matched with the free space impedance at normal incidence.

## VIII. CONCLUSIONS

The absorption properties of metamaterial absorbers have been analyzed by resorting to a simple equivalent transmission line circuit. The closed-form expression of the input impedance of the printed structure has been derived as a function of the loss component of the dielectric substrate and of the FSS capacitance. It is shown that the real part of the input impedance, which determines the amount of loss, is composed by three loss terms: one due to the classical ohmic loss mechanism while the other two terms are determined by dielectric losses.

The effect of FSS element geometry and substrate loss component on the reflection loss is analyzed through some practical examples. It is shown that highly capacitive elements, keeping the substrate parameters fixed, need higher substrate losses with respect to low capacitive elements to perform perfect absorption but they lead to the maximization of the absorption bandwidth. The bandwidth where the metamaterial absorber performs absorption/emission can be tailored by adequately tuning the geometrical and electrical parameters of the structure.

The angular stability of the absorbing structure is finally addressed presenting simple design criteria for achieving good absorption also at very wide incident angles.

A physical interpretation of the absorption phenomenon and simple analytical guidelines to exploit all the degrees of freedom have been provided.